\documentclass[%
 reprint,
preprintnumbers,
nofootinbib,
 amsmath,amssymb,
 aps,
]{revtex4-2}
\usepackage{amsmath}
\usepackage{float}
\usepackage{tabularx}
\usepackage{graphicx}
\usepackage{braket}
\usepackage{makecell}
\usepackage{tikz}
\usepackage{xspace}
\usepackage{orcidlink}
\usepackage{multirow}
\usepackage{enumitem}
\setlist[enumerate]{label*=\arabic*.}
\usepackage{tikz}

\usetikzlibrary{shapes.geometric, arrows}

\tikzstyle{startstop} = [rectangle, rounded corners, minimum width=3cm, minimum height=1cm,text centered, draw=black, fill=red!30]
\tikzstyle{io} = [trapezium, trapezium left angle=70, trapezium right angle=110, text width=3cm, minimum height=1cm, text centered, draw=black, fill=blue!30]
\tikzstyle{process} = [draw,rectangle, text width=5cm, minimum height=1cm, text centered, draw=black, fill=orange!30]
\tikzstyle{process2} = [draw,rectangle, text width=1cm, minimum height=1cm, text centered, draw=black, fill=orange!30]
\tikzstyle{decision} = [diamond, text width=2cm, text badly centered, draw=black, fill=green!30, inner sep=0pt]
\tikzstyle{arrow} = [thick,->,>=stealth]

\begin{document}

\title{Impact of Population III stars on the astrophysical gravitational-wave background}

\author{Nikolaos~Kouvatsos\,\orcidlink{0000-0002-5497-3401}}
\email{nikolaos.kouvatsos@kcl.ac.uk}
\affiliation{Theoretical Particle Physics and Cosmology Group, Physics Department, King's College London, University of London, Strand, London WC2R 2LS, UK}

\author{Mairi~Sakellariadou\,\orcidlink{0000-0002-2715-1517}}
\email{mairi.sakellariadou@kcl.ac.uk}
\affiliation{Theoretical Particle Physics and Cosmology Group, Physics Department, King's College London, University of London, Strand, London WC2R 2LS, UK}

\date{\today}
\preprint{KCL-PH-TH/2024-17}

\begin{abstract}
   We probe the astrophysical gravitational-wave background resulting from compact binary coalescences, focusing on Population III binary black holes. We exploit results of state-of-the-art simulations on the evolution of Population I-II and III binaries, considering a variety of initial condition and star formation rate models for the latter. 
   The contribution from Population III binary black holes is found to be very small, with no effect on  the gravitational-wave spectrum. 
   A network of third-generation detectors will detect easier individual Population III binaries, due to their significantly higher masses, hence decreasing even further their residual contribution.
\end{abstract}

\maketitle

\section{Introduction}

During its first three observing runs, the LIGO-Virgo-KAGRA (LVK) Collaboration has detected almost a hundred gravitational wave (GW) signals from compact binary coalescences (CBCs)~\cite{PhysRevX.9.031040, PhysRevX.11.021053, 2021arXiv211103606T}, while at the end of the ongoing O4 run this number is expected to increase considerably~\cite{KAGRA:2013rdx}. This multitude of detections allow us to continuously refine our understanding of the stellar population of compact binaries and constrain astrophysical models. Apart from the transient signals, there is a continuous effort to search for the gravitational-wave background, a random signal made up of the superposition of numerous GWs throughout the history of the Universe~\cite{1997rggr.conf..373A, Christensen_2019}.

While a gravitational-wave background may result from various cosmological mechanisms~\cite{Caprini_2018}, the astrophysical GW background (AGWB) made up from CBCs mergers is expected to be the dominant contributor~\cite{Regimbau:2008nj,Tania_Regimbau_2011}.  To estimate the AGWB, only binaries produced by Population I-II stars are usually considered. There is, however, an older population of stars - typically addressed in the literature as Population III stars - which could result in a sizeable contribution to the AGWB~\cite{refId0, 10.1093/mnras/stw1431, Perigois_2020, Martinovic_2022}.

The increasing number of studies focusing on Population III stars in recent years~\cite{Belczynski_2004, 10.1093/mnras/stu1022, 10.1093/mnrasl/slw074, 10.1093/mnras/stx757, 10.1093/mnras/stx1759, 10.1093/mnras/staa2511, Tanikawa_2021, 10.1093/mnras/stab1421, 10.1093/mnrasl/slaa191, 10.1093/ptep/ptaa176, 10.1093/mnrasl/slab032, 10.1093/pasj/psac010, Filippo_et_al, Costa_et_al, tanikawa2024contribution}, combined  with the updated information gained on Population I \& II stars from the LVK Collaboration detections, motivates us to revisit the AGWB from both Population I-II and Population III binaries. We compute the expected total AGWB from Population I-II \& III binaries for a network of 3G detectors, as well as the unresolved AGWB made up of binaries that are not individually detectable. We show that the contribution of GW signals from Population III mergers to the total AGWB is negligible, as already found in previous studies~\cite{Martinovic_2022}. 
We then show that since compact binaries from Population III stars are expected to be much more massive than those of Population I-II stars, hence easier to be detected individually in the astrophysical GW \textit{foreground}, their contribution to the AGWB is completely lost after foreground subtraction.

This paper is organised as follows: In Section~II,  we revisit the standard formalism to study the AGWB, focusing on an approach with  catalogues of sources. We then  describe the simulated population models and the corresponding compact binary parameters. In Section~\ref{unresolved_background}, we outline the process to remove the astrophysical GW foreground detected by a 3G detector network. In Section~\ref{results} we present our findings for the total and unresolved AGWB. We also discuss statistics for Population I-II and III binaries in terms of the signal-to-noise ratio. In Section~V, we summarise our conclusions and briefly comment on the differences between our and related previous studies.

\section{Astrophysical gravitational-wave background}

We briefly outline the formalism to estimate the AGWB. We then summarise our method to build catalogues of compact binaries and highlight our selected binary parameters. 

\subsection{Formalism}

We describe the AGWB by the normalised GW energy spectrum~\cite{Allen_Romano}
\begin{equation}
    \Omega_{\rm GW}(f)=\frac{1}{c^2\rho_{\rm c}}\frac{{\rm d}\rho_{\rm GW}}{{\rm d}\ln f}=\frac{f}{c^3\rho_{\rm c}}F(f),
\end{equation}
where $\rho_{\rm c}=3H_0^2/(8\pi G)$ is the critical energy density, with $H_0$ the Hubble's constant and $G$ the Newton's gravitational constant, $c$ the speed of light,  and $f$ the GW frequency in the detector frame. The integrated flux density~\cite{Regimbau:2008},
\begin{equation}
    F(f)=\int_0^{z_{\rm max}}\frac{R(z)}{4\pi d_c^2(z)}\frac{{\rm d}E_{\rm GW}}{{\rm d}f_{\rm s}}\bigg\vert_{f_{\rm s}}{\rm d}z,
    \label{integrated_flux_density}
\end{equation}
depends on the merger rate, $R(z)$, the comoving distance, $d_{\rm c}(z)$, and the GW energy spectrum in the source frame, ${\rm d}E_{\rm GW}/{\rm d}f_{\rm s}$, with $f_{\rm s}=f(1+z)$ the GW frequency in the source frame.
We express the merger rate $R(z)$ in terms of the merger rate per unit comoving volume, $R_{\rm V}(z)$, as
\begin{equation}
    R(z)=\frac{R_{\rm V}(z)}{1+z}\frac{{\rm d}V}{{\rm d}z}=\frac{4\pi c}{H_0}\frac{R_{\rm V}(z)d_{\rm c}^2(z)}{(1+z)E(z)},
    \label{R}
\end{equation}
where $E(z)=\sqrt{\Omega_{\rm m}(1+z)^3+\Omega_\Lambda}$. We adopt a flat $\Lambda$CDM model with $\Omega_{\rm m}=0.31$, $\Omega_\Lambda=0.69$ and $H_0=67.66 {\rm km}/{\rm s}/{\rm Mpc}$ \cite{Planck18}. The $(1+z)$ factor accounts for cosmic expansion.

The GW energy spectrum for a single source reads  \cite{Phinney}
\begin{equation}
    \frac{{\rm d}E_{\rm GW}}{{\rm d}f_{\rm s}}\Big\vert_{f_{\rm s}}=\frac{2\pi^2 d_{\rm c}^2c^3}{G}f_{\rm s}^2\Big(\big\vert\Tilde{h}_+(f_{\rm s})\big\vert^2+\big\vert\Tilde{h}_\times(f_{\rm s})\big\vert^2\Big),
    \label{dEdf}
\end{equation}
where $\Tilde{h}_+$, $\Tilde{h}_\times$ are the Fourier transforms of the two polarisation modes of the gravitational waveform in the frequency domain.

Given a catalogue of sources, the GW energy spectrum, Eq.~(\ref{dEdf}), can be computed for each source individually. The integrated flux density, Eq. (\ref{integrated_flux_density}), can then be replaced by a sum over all sources \cite{PhysRevLett.118.151105}:
\begin{equation}
    F(f) = \frac{1}{T}\frac{\pi c^3}{2G}f^2\sum_{i=1}^N\Big(\big\vert\Tilde{h}_+^i(f)\big\vert^2+\big\vert\Tilde{h}_\times^i(f)\big\vert^2\Big),
    \label{new_F}
\end{equation}
where $\Tilde{h}_+$, $\Tilde{h}_\times$ are now taken in the detector frame, where $i$ denotes a particular source. The total number of sources, $N$, is related to the total time of observation, $T$, as:
\begin{equation}
    N = T\int_0^{z_{\rm max}} \frac{R_{\rm V}(z)}{1+z}\frac{{\rm d}V}{{\rm d}z}{\rm d}z = T\int_0^{z_{\rm max}}\frac{4\pi c}{H_0}\frac{R_{\rm V}(z)d_{\rm c}^2(z)}{(1+z)E(z)}{\rm d}z.
    \label{N_T}
\end{equation}

In what follows, we use catalogues of simulated GW sources, which have been stored in redshift bins. We thus perform a discrete summation to calculate $N$, instead of computing the integral expression given in Eq.~(\ref{N_T}). The number of sources at each redshift bin is then given by
\begin{equation}
    N_i = T\frac{4\pi c}{H_0}\frac{R_{\rm V}(z_i)d_{\rm c}^2(z_i)}{(1+z_i)E(z_i)}\Delta z_i,
    \label{N_i}
\end{equation}
with total number of sources 
\begin{equation}
    N = \sum_i^{z_{{\rm max}}^i}N_i.
\end{equation}
We opt for a redshift resolution  $\Delta z_i = 0.1, \forall i$ and  fix $N = 10^5$ for statistical convergence (as in Ref.~\cite{PhysRevD.108.064040}).


\subsection{Catalogues}\label{pop_models}

We use binaries simulated and evolved utilising the {\it Stellar EVolution for N-body} (\texttt{SEVN}) binary population synthesis code~\cite{Spera_2019, Mapelli_2020, Iorio_et_al}. We make a distinction between Population I-II and Population III binaries. For the former, we pick the \textit{Fiducial} model from Ref.~\cite{Iorio_et_al}, for three distinct classes of compact binaries: binary black holes (BBHs), black hole-neutron stars (BHNSs), and binary neutron stars (BNSs). For the latter, we use results from Ref.~\cite{Costa_et_al} (in which only BBHs have been produced\footnote{Past studies on supernova remnants from Population III have focused on black holes, which is why we choose to neglect the contributions to the $\Omega_{\rm GW}$ from the BHNS \& BNS subpopulations as subdominant.  We refer the reader to Ref.~\cite{Belczynski_2004, 10.1093/mnras/stu1022, 10.1093/mnrasl/slw074, 10.1093/mnras/stx757, 10.1093/mnras/stx1759, 10.1093/mnras/staa2511, Tanikawa_2021, 10.1093/mnras/stab1421, 10.1093/mnrasl/slaa191, 10.1093/ptep/ptaa176, 10.1093/mnrasl/slab032, 10.1093/pasj/psac010, Filippo_et_al, Costa_et_al, tanikawa2024contribution}.}), considering 11 distinct models, each assuming different initial conditions (ICs) (denoted by \textit{LOG1}, \textit{LOG2}, \textit{LOG3}, \textit{LOG4}, \textit{LOG5}, \textit{KRO1}, \textit{KRO5}, \textit{LAR1}, \textit{LAR5}, \textit{TOP1}, \textit{TOP5}).

 The \texttt{SEVN} binary population synthesis code, provides 3 of the binary parameters, namely the two compact binary masses and the eccentricity of the orbit. 

The Star Formation Rate (SFR) is expected to be different for Population I-II and Population III stars, based on their different metallicity. For the former, we consider the metallicity-dependent Madau-Fragos SFR model~\cite{Madau_2017}, with metallicity bins: $Z\in\{0.0001, 0.0002, 0.0004, 0.0006, 0.0008, 0.001, 0.002, 0.004$, $0.006, 0.008, 0.01, 0.014, 0.017, 0.02, 0.03\}$~\cite{Filippo_et_al}. For the latter, we consider 4 different SFR models~\cite{Hartwig_2022, 10.1093/mnras/stz1529, 10.1093/mnras/staa2143, 10.1093/mnras/staa139} (denoted by \textit{H22}, \textit{J19}, \textit{LB20}, \textit{SW20} in Ref.~\cite{Filippo_et_al}), all with the same metallicity $Z=10^{-11}$~\cite{Filippo_et_al}.

The high number of considered IC and SFR models aims to account for the high uncertainty that characterises the properties of Population III stars. For a brief description of the different models, see \hyperref[appendix_a]{Appendix A}.

Once we have the Star Formation Rate (SFR), we use the (publicly available) code \texttt{CosmoRate}~\cite{CosmoRate_1, CosmoRate_2} to get -- for a given redshift bin -- the Merger Rate Density (MRD) and a set of binaries, from which we randomly select $N_i$, Eq. (\ref{N_i}),  to produce a catalogue.

The aforementioned steps to produce  a catalogue are summarised in the flowchart,  Fig.~\ref{flowchart}.

\tikzset{every picture/.style={line width=0.75pt}} 

\begin{figure}
\begin{tikzpicture}[x=0.75pt,y=0.75pt,yscale=-1,xscale=1]

\draw  [fill={rgb, 255:red, 114; green, 134; blue, 160 }  ,fill opacity=1 ] (65,78.9) .. controls (65,76.19) and (67.19,74) .. (69.9,74) -- (106.1,74) .. controls (108.81,74) and (111,76.19) .. (111,78.9) -- (111,97.1) .. controls (111,99.81) and (108.81,102) .. (106.1,102) -- (69.9,102) .. controls (67.19,102) and (65,99.81) .. (65,97.1) -- cycle ;
\draw  [fill={rgb, 255:red, 205; green, 231; blue, 176 }  ,fill opacity=1 ] (56.56,139) -- (137,139) -- (118.44,179) -- (38,179) -- cycle ;
\draw  [fill={rgb, 255:red, 163; green, 191; blue, 168 }  ,fill opacity=1 ] (26,211) -- (151,211) -- (151,259) -- (26,259) -- cycle ;
\draw  [fill={rgb, 255:red, 114; green, 134; blue, 160 }  ,fill opacity=1 ] (205,223) .. controls (205,219.13) and (208.13,216) .. (212,216) -- (268,216) .. controls (271.87,216) and (275,219.13) .. (275,223) -- (275,249) .. controls (275,252.87) and (271.87,256) .. (268,256) -- (212,256) .. controls (208.13,256) and (205,252.87) .. (205,249) -- cycle ;
\draw  [fill={rgb, 255:red, 205; green, 231; blue, 176 }  ,fill opacity=1 ] (105.94,315) -- (227,315) -- (199.06,355) -- (78,355) -- cycle ;
\draw  [fill={rgb, 255:red, 163; green, 191; blue, 168 }  ,fill opacity=1 ] (48,412) -- (118,412) -- (118,452) -- (48,452) -- cycle ;
\draw  [fill={rgb, 255:red, 163; green, 191; blue, 168 }  ,fill opacity=1 ] (171,397) -- (290,397) -- (290,470) -- (171,470) -- cycle ;
\draw  [fill={rgb, 255:red, 190; green, 110; blue, 70 }  ,fill opacity=1 ] (88,522) .. controls (88,517.58) and (91.58,514) .. (96,514) -- (223,514) .. controls (227.42,514) and (231,517.58) .. (231,522) -- (231,546) .. controls (231,550.42) and (227.42,554) .. (223,554) -- (96,554) .. controls (91.58,554) and (88,550.42) .. (88,546) -- cycle ;
\draw    (90,102) -- (90,138) ;
\draw [shift={(90,137.3)}, rotate = 270] [color={rgb, 255:red, 0; green, 0; blue, 0 }  ][line width=0.75]    (10.93,-3.29) .. controls (6.95,-1.4) and (3.31,-0.3) .. (0,0) .. controls (3.31,0.3) and (6.95,1.4) .. (10.93,3.29)   ;
\draw    (89,179) -- (89,210) ;
\draw [shift={(89,208.4)}, rotate = 270] [color={rgb, 255:red, 0; green, 0; blue, 0 }  ][line width=0.75]    (10.93,-3.29) .. controls (6.95,-1.4) and (3.31,-0.3) .. (0,0) .. controls (3.31,0.3) and (6.95,1.4) .. (10.93,3.29)   ;
\draw    (86,259) -- (86,280) ;
\draw    (242,256) -- (242,280) ;
\draw    (86,280) -- (242,280) ;
\draw    (154,355) -- (154,376) ;
\draw    (232,470) -- (232,493) ;
\draw    (83,452) -- (83,493) ;
\draw    (83,493) -- (232,493) ;
\draw    (154,376) -- (231,376) ;
\draw    (154,376) -- (83,376) ;
\draw    (156,280) -- (156,312) ;
\draw [shift={(156,313.2)}, rotate = 270] [color={rgb, 255:red, 0; green, 0; blue, 0 }  ][line width=0.75]    (10.93,-3.29) .. controls (6.95,-1.4) and (3.31,-0.3) .. (0,0) .. controls (3.31,0.3) and (6.95,1.4) .. (10.93,3.29)   ;
\draw    (83,376) -- (83,409) ;
\draw [shift={(83,410)}, rotate = 270] [color={rgb, 255:red, 0; green, 0; blue, 0 }  ][line width=0.75]    (10.93,-3.29) .. controls (6.95,-1.4) and (3.31,-0.3) .. (0,0) .. controls (3.31,0.3) and (6.95,1.4) .. (10.93,3.29)   ;
\draw    (231,376) -- (231,395) ;
\draw [shift={(231,395)}, rotate = 270] [color={rgb, 255:red, 0; green, 0; blue, 0 }  ][line width=0.75]    (10.93,-3.29) .. controls (6.95,-1.4) and (3.31,-0.3) .. (0,0) .. controls (3.31,0.3) and (6.95,1.4) .. (10.93,3.29)   ;
\draw    (157.5,493) -- (157.04,514) ;
\draw [shift={(157,512)}, rotate = 271.25] [color={rgb, 255:red, 0; green, 0; blue, 0 }  ][line width=0.75]    (10.93,-3.29) .. controls (6.95,-1.4) and (3.31,-0.3) .. (0,0) .. controls (3.31,0.3) and (6.95,1.4) .. (10.93,3.29)   ;

\draw (69.2,81.7) node [anchor=north west][inner sep=0.75pt]  [color={rgb, 255:red, 0; green, 0; blue, 0 }  ,opacity=1 ] [align=left] {1. ICs};
\draw (63,152.7) node [anchor=north west][inner sep=0.75pt]   [align=left] {2. \texttt{SEVN}};
\draw (42,221) node [anchor=north west][inner sep=0.75pt]   [align=left] { \ \ \ \ 3a. List of\\evolved binaries};
\draw (214,229.5) node [anchor=north west][inner sep=0.75pt]   [align=left] {3b. SFR};
\draw (111,329) node [anchor=north west][inner sep=0.75pt]   [align=left] {4. \texttt{CosmoRate}};
\draw (54,425.6) node [anchor=north west][inner sep=0.75pt]   [align=left] {5a. MRD};
\draw (185,412.5) node [anchor=north west][inner sep=0.75pt]   [align=left] { \ \ \ \ 5b. Set of\\binaries at each\\ \ \ \ redshift bin};
\draw (101.3,527.5) node [anchor=north west][inner sep=0.75pt]   [align=left] {6. Binary catalogues};
\end{tikzpicture}
\caption{Flowchart summarising the steps followed to construct the binary catalogues, for a model with given initial conditions (ICs) and star formation rate (SFR). Steps 1-3a have already been performed in Ref.~\cite{Iorio_et_al, Costa_et_al}. The list of evolved binaries (step 3a) can be found in Ref.~\cite{Zenodo_Iorio, Zenodo_Costa}.}
\label{flowchart}
\end{figure}

As mentioned previously,  \texttt{SEVN} provides only the compact binary masses in the source frame, $m_{\rm 1(source)}$ and $m_{\rm 2(source)}$ (where 1 stands for the primary and 2 for the secondary compact object), and the eccentricity, $e$. We have to choose 8 additional parameters:  2 intrinsic parameters (spins) and 6 extrinsic ones (polarisation angle,
coalescence phase, right ascension, declination, inclination angle, and coalescence time). We sample as follows (see, Table~\ref{table_dist}):
\begin{itemize}
    \item The spins in the $z$-direction, $\chi_{1(z)}$ and $\chi_{2(z)}$, are sampled from uniform distributions in the range $[-0.75, 0.75]$ for black holes, and $[-0.05, 0.05]$ for neutron stars~\cite{Gupta:2023evt}. We set the spins in the $x$- and $y$-direction to 0.
    \item The polarisation angle, $\psi$, coalescence phase, $\phi_{\rm c}$, right ascension, $\alpha$, declination, $\delta$, and inclination angle, $\iota$ are sampled from uniform distributions in $\psi$, $\alpha$, $\phi_{\rm c}$, ${\rm cos}(\iota)$, ${\rm cos}(\delta + \pi/2)$.
    \item The coalescence time in the detector data segment, $t_{\rm c}$, is set at 0 for all signals.
\end{itemize}

\begin{table}
\centering
\begin{tabular}{ |c|c|c| }
 \hline
 & \multicolumn{2}{|c|}{\textbf{Sampling distributions}} \\
 \hline
\textbf{Parameter} & \textbf{Black holes} & \textbf{Neutron stars}\\
 \hline
$m_{\rm 1(source)}$, $m_{\rm 2(source)}$, $e$, $z$ & \multicolumn{2}{|c|}{\texttt{SEVN} \& \texttt{CosmoRate}}\\
\hline
$\chi_{1(x)}$, $\chi_{1(y)}$, $\chi_{2(x)}$, $\chi_{2(y)}$, $t_{\rm c}$ & \multicolumn{2}{|c|}{0}\\
\hline
\multirow{2}{*}{$\chi_{1(z)}$, $\chi_{2(z)}$} & Uniform in & Uniform in\\
& $[-0.75, 0.75]$ & $[-0.05, 0.05]$\\
\hline
$\psi$, $\alpha$, $\phi_{\rm c}$ & \multicolumn{2}{|c|}{Uniform in $[0, 2\pi]$}\\
\hline
${\rm cos}(\iota)$, ${\rm cos} (\delta + \pi/2)$ & \multicolumn{2}{|c|}{Uniform in $[-1, 1]$}\\
 \hline
\end{tabular}
\caption{Summary of the distributions we use to sample in all compact binary parameters.}
\label{table_dist}
\end{table}

As for the gravitational waveform we use to model the signals, we have opted for \texttt{IMRPhenomXHM}~\cite{IMRPHENOMXHM}. This choice has been motivated by a preference for probing the harmonic modes of the signals to compute the AGWB (as considered in Ref.~\cite{Perigois_2020}), rather than precession.

\section{Unresolved AGWB}\label{unresolved_background}
An AGWB present in the data will be dominated by a foreground of loud detected signals; removing them from the dataset would allow to estimate the AGWB truly composed of individually unresolvable signals. 

The unresolved contribution to the $\Omega_{\rm GW}$ is made up of all signals below some SNR threshold. For the $i-$th binary, we define the network optimal SNR as
\begin{equation}
    \rho_i^2 = \sum_{I=1}^M4\int_{f_{\rm min}}^{f_{\rm max}}\frac{\big\vert\Tilde{H}_I^i(f)\big\vert^2}{P_I(f)}{\rm d}f,
\end{equation}
where
\begin{equation}
\begin{aligned}
    \Tilde{H}_I^i(f) = {} & F_{\rm lp}^{I,i}(f,\alpha_i,\delta_i)\big[F_+^{I,i}(f,\alpha_i,\delta_i,\psi_i)\Tilde{h}_+^i(f)\\
    &+F_\times^{I,i}(f,\alpha_i,\delta_i,\psi_i)\Tilde{h}_\times^i(f)\big]
\end{aligned}
\end{equation}
is the detector response, $P_I$ the power spectral density of detector $I$, and M the total number of considered detectors. We denote by $F_{\rm lp}$  the location-phase factor, and by $F_+$, $F_\times$  the antenna pattern functions.

We consider a three 3G detector network: an L-shaped Cosmic Explorer (CE) detector with 20km-long arms and post-merger optimised PSD located at the site of LIGO Hanford, an L-shaped CE detector with 40km-long arms and compact-binary optimised PSD located at the site of LIGO Livingston~\cite{Srivastava_2022}, and a triangular (that is, composed of three distinct interferometers) Einstein Telescope (ET) detector with 10km-long arms located at the site of Virgo~\cite{Sathyaprakash_2012}. We set the minimum frequency at $f_{\rm min}=5$ Hz and the maximum frequency at $f_{\rm max}=1024$ Hz~\cite{PhysRevD.105.104016}, as contributions to the SNR outside this frequency range are negligible. To compute the SNR we use the \texttt{GWBENCH} package~\cite{Borhanian_2021}.

We compute the optimal SNR of every signal in our catalogue. The calculation of the unresolved AGWB follows the same procedure as outlined for the total AGWB, with the sole difference that this time we only consider the signals in our catalogues that pass the SNR threshold. We consider two choices for the SNR threshold: $\rho_{\rm thr}$ = 12 (conservative), and  $\rho_{\rm thr}$ = 8.

\section{Results}\label{results}

In what follows, we present the total AGWB produced separately by mergers of Population I-II stars and Population III stars, and discuss the characteristics of the latter versus the former. We then discuss the difference in the unresolved AGWB for Population I-II versus Population III, noting that this depends on the detector sensitivities. In our study, we have used a three 3G detector network and fixed the SNR threshold.
  
\subsection{Total background}\label{total_background}

We plot in Fig. \ref{Pop_I_II_III_tot} the total $\Omega_{\rm GW}$ from mergers of all three binary classes (BBHs, BHNSs, and BNSs) produced by Population I-II stars, as well as the individual contributions. We also plot the $\Omega_{\rm GW}$ resulting from Population III binaries, considering the most optimistic (\textit{J19}, \textit{LAR1}) and most pessimistic (\textit{SW20}, \textit{TOP5}) cases (we refer the reader to {\hyperref[appendix]{Appendix B}, where we present results for all Population III cases). We also show the total AGWB from Population I-II, and III binaries, where for the latter we selected the most optimistic model.

We observe that for Population I-II, the BHNS contribution is the weakest. The AGWB seems to be dominated by the BBH mergers in the whole frequency range except for the highest frequencies, above $f\sim900$ Hz, where the BNS contribution prevails. The contribution from Population III binaries peaks at $\sim10^{-13}$ or $\sim6\times10^{-11}$ depending on the assumed SFR and IC model. In all cases, we observe a plateau which in the most optimistic scenario is in the region $f\in[20, 200]$ Hz. Clearly, even in this most optimistic case, the resulting increase in the total AGWB after including BBHs from Population III stars is only minimal, hence insufficient to lead to significant deviation from the expected $\propto f^{2/3}$ spectrum~\cite{Phinney}.
Considering the contribution from Population I-II and III binaries, the $\Omega_{\rm GW}$ peaks just above $10^{-9}$. 

We next investigate the AGWB from Population I-II and Population III binaries  when considering only unresolved sources.

\begin{figure}
\centering
\includegraphics[width=\linewidth]{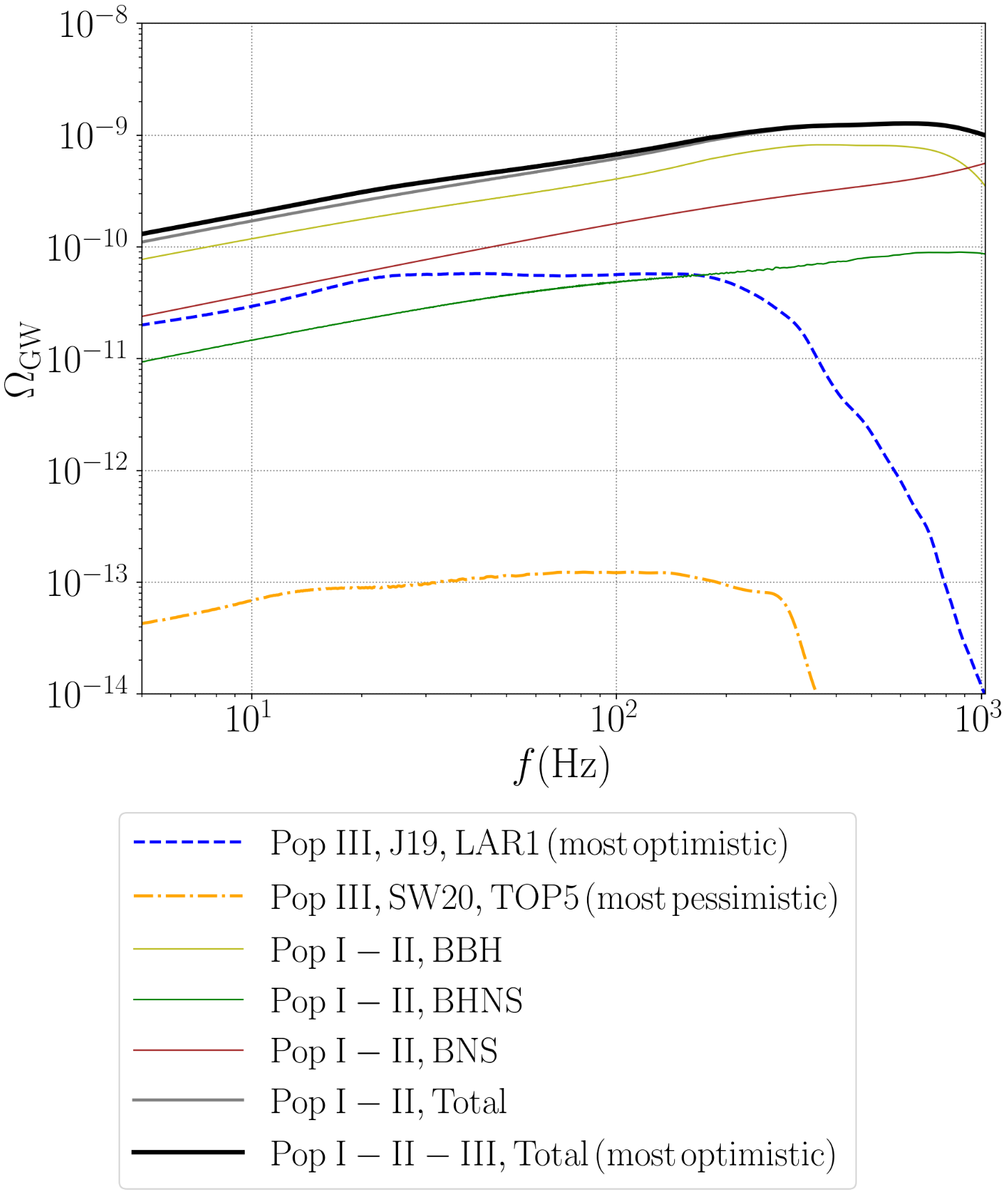}
\caption{The AGWB resulting from mergers of BBHs, BHNSs, and BNSs produced by Population I-II stars, as well as their total. Including the BBH mergers from Population III stars has a negligible impact even when considering the most optimistic model. The AGWB for the most pessimistic Population III model is also provided for comparison.}
\label{Pop_I_II_III_tot}
\end{figure}

\subsection{Unresolved background}\label{unresolved_GWB}
We plot in Fig.~\ref{unresolved_Omega} the unresolved AGWB for our two choices of $\rho_{\rm thr}$. First, we consider the Population I-II contribution. Imposing an SNR threshold of 12 seems to drop the $\Omega_{\rm GW}$ by $\sim$ 1 order of magnitude in the whole frequency range except for the highest frequencies, where the BNS mergers contribute most. This is expected, as GW signals from BNSs are weaker than those from BBHs and dominate after foreground subtraction.

We next consider Population III for the most optimistic case (\textit{J19}, \textit{LAR1}). One would expect BBH mergers from Population III stars to be overall harder to detect than those of Population I-II stars, given that the merger rate density for Population III binaries peaks at higher redshifts (see \hyperref[appendix]{Appendix B} for a detailed figure of the computed MRD). We observe, however, that the corresponding unresolved $\Omega_{\rm GW}$, for an SNR threshold of 12, lies $\sim2$ orders of magnitude below the total $\Omega_{\rm GW}$ for $f\leq 30$ Hz, where it follows the $\propto f^{2/3}$ dependence. Notice that $\Omega_{\rm GW}$ has lost it characteristic plateau, dropping abruptly for frequencies above $f\sim 30$ Hz. Such frequencies correspond to BBHs with very low redshifts, thus high SNRs, therefore removed by the foreground subtraction.

To further quantify the difference in the foreground subtraction between Population I-II and III binaries, we show in Fig. \ref{SNR_hist} a histogram of the network optimal SNR's for BBH mergers from Population I-II \& III (\textit{J19}, \textit{LAR1}) binaries. The SNR distribution for Population III appears noticeably shifted towards higher SNR's compared to the same distribution for Population I-II, with a mean SNR of 38.0 instead of 27.3, respectively. It is, therefore, sensible that in the case of Population III the deviation of the unresolved AGWB from the total one is greater.

Note that these results assume a perfect foreground subtraction. Even though this is not the case~\cite{PhysRevD.108.064040, 2024arXiv240100984S}, we are not considering any errors that would decrease the efficiency of the subtraction, since the contribution from Population III binaries to the total AGWB is quite small, and lost after foreground subtraction.

\begin{figure}
\centering
\includegraphics[width=\linewidth]{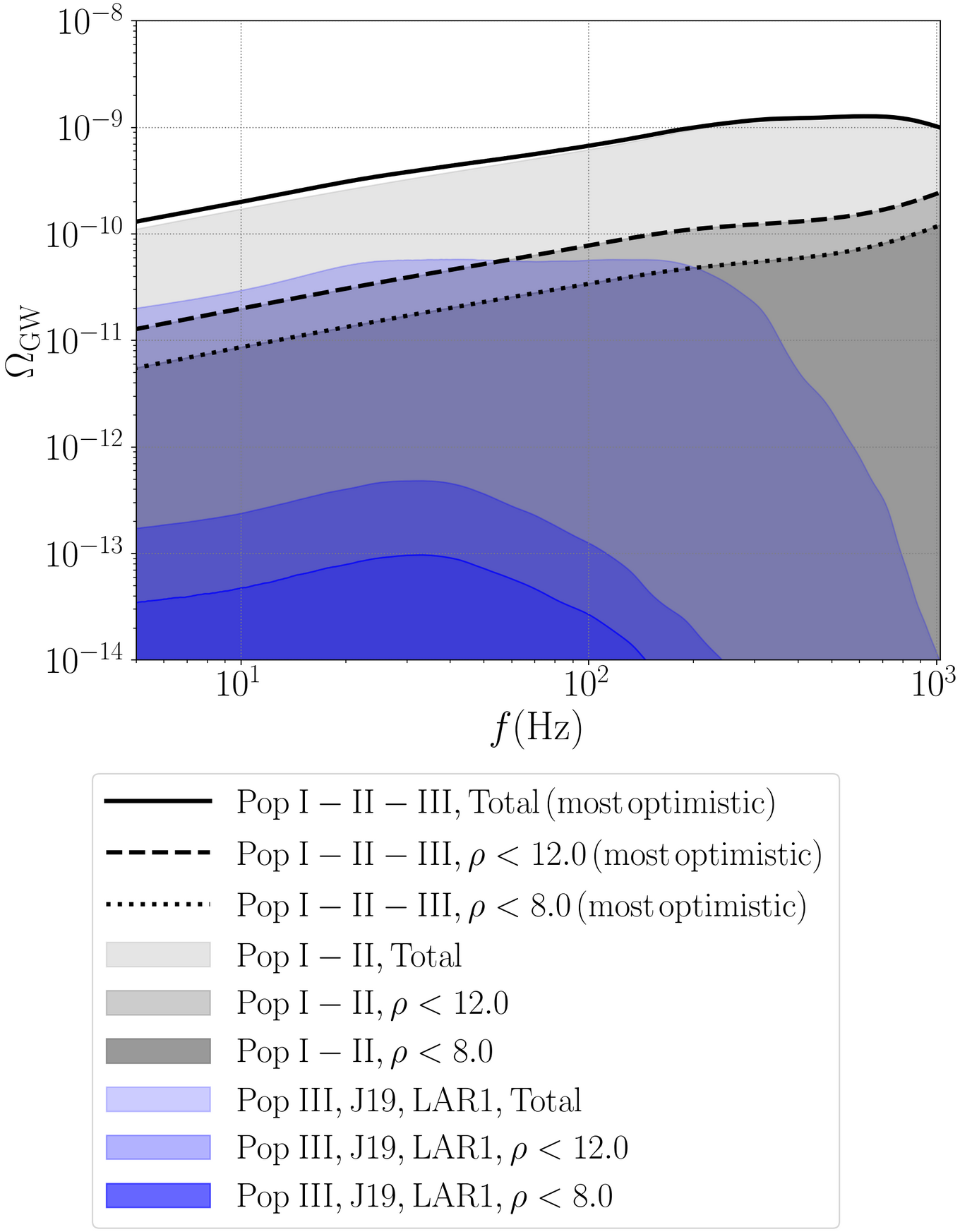}
\caption{The unresolved AGWB resulting from mergers of binaries produced by Population I-II \& III stars, for an SNR threshold $\rho_{\rm thr}\in\{8,12\}$. The AGWB before the foreground subtraction is also plotted for comparison. Removing the foreground of unresolvable sources from our catalogue seems to reduce the AGWB more in the case of Population III binaries. Finally, we provide the total AGWB from all three Populations before and after the foreground subtraction.}
\label{unresolved_Omega}
\end{figure}

\begin{figure}
\centering
\includegraphics[width=\linewidth]{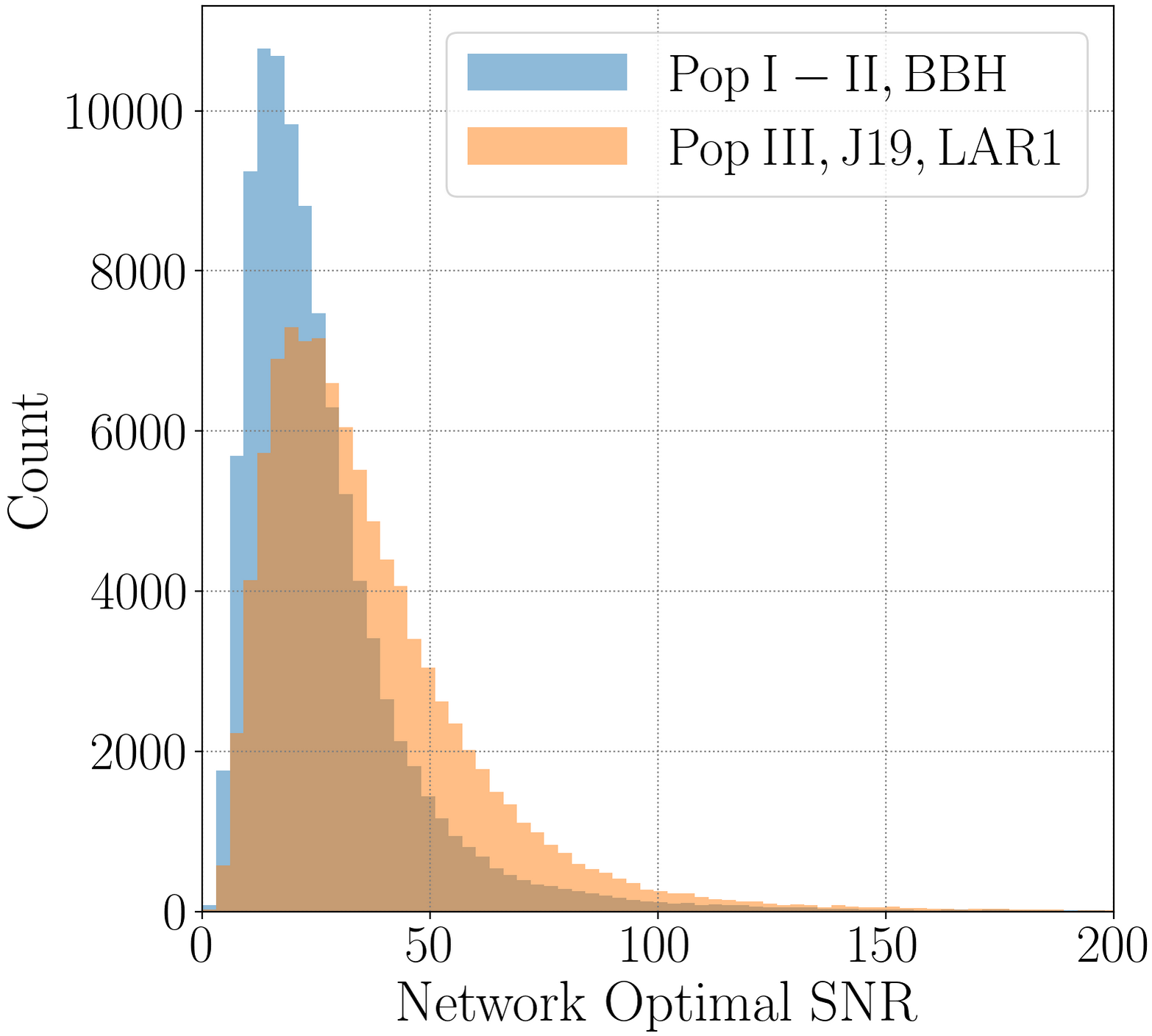}
\caption{Histogram with the network optimal SNRs of all BBHs for Population I-II \& Population III (\textit{J19}, \textit{LAR1}). Despite the higher redshifts typically associated with Population III BBHs compared to Population I-II, the distribution of the former is shifted towards higher SNRs due their significantly higher masses.}
\label{SNR_hist}
\end{figure}

\section{Discussion \& Conclusions}

We have studied the contribution from Population III compact binaries mergers to the astrophysical gravitational-wave background and compared to that from Population I-II.  For Population I-II binaries, we have considered all three types of mergers (BBH, BNS, BHNS). For Population III, we have considered a variety of models with different initial conditions for the binary population synthesis code and different star formation rate.

Our analysis has determined that including GW signals from Population III binaries has a very small impact on the total $\Omega_{\rm GW}$, as already shown in Ref.~\cite{Perigois_2020}. Our study, however, has yielded results that contradict expectations on the unresolved AGWB. Specifically, we find Population III binaries to be on average easier to detect than Population I-II binaries, regardless of the assumed initial conditions and star formation rate of the former. For a given SNR threshold, foreground subtraction has a greater impact in the case of Population III binaries, resulting in their contribution to the total AGWB being lost. Hence the AGWB is characterised and distinguished from the cosmological GWB by its $\propto f^{2/3}$ spectrum. Loud individual GW signals from Population III binaries could be present in the data and identifiable from their higher redshifts compared to Population I-II binaries.

Let us note that the difference in our results with respect to previous ones~\cite{Perigois_2020, Martinovic_2022} is due mainly to the different merger rates and binary masses of the considered catalogues. To build our catalogues, we have employed results from the \texttt{SEVN} binary population synthesis code, whereas the study of Ref.~\cite{Perigois_2020, Martinovic_2022} is based on simulations with \texttt{StarTrack}~\cite{StarTrack}. We refer the reader to models \textit{M10} and \textit{FS1} for Population I-II and III, respectively, of Ref.~\cite{10.1093/mnras/stx1759}.
In our models the MRD peaks at   $z\simeq4.5$ and $z\geq7.5$ for Population I-II and Population III, respectively, whereas for the \texttt{StarTrack} models the MRD peaks correspondingly at $z\simeq2$ and $z\simeq12$. For our models, the average BBH total source mass is $M_{\rm source}=16.8 M_\odot$ for Population I-II, and ranges from $M_{\rm source}=49.6 M_\odot$ (\textit{H22}, \textit{KRO1}) to $M_{\rm source}=64.0 M_\odot$ for Population III binaries (see \hyperref[appendix]{Appendix B} for a detailed figure of the total source mass distribution), whereas for the \texttt{StarTrack} models the corresponding masses (for mergers within $z<2$) are $M_{\rm source}=29.7M_\odot$ and $M_{\rm source}=63.4 M_\odot$. Finally, we note that the BHNS and BNS waveform adopted in Ref.~\cite{Perigois_2020, Martinovic_2022} considered only the inspiral phase.
 
\begin{figure*}
\centering
\includegraphics[scale=0.28]{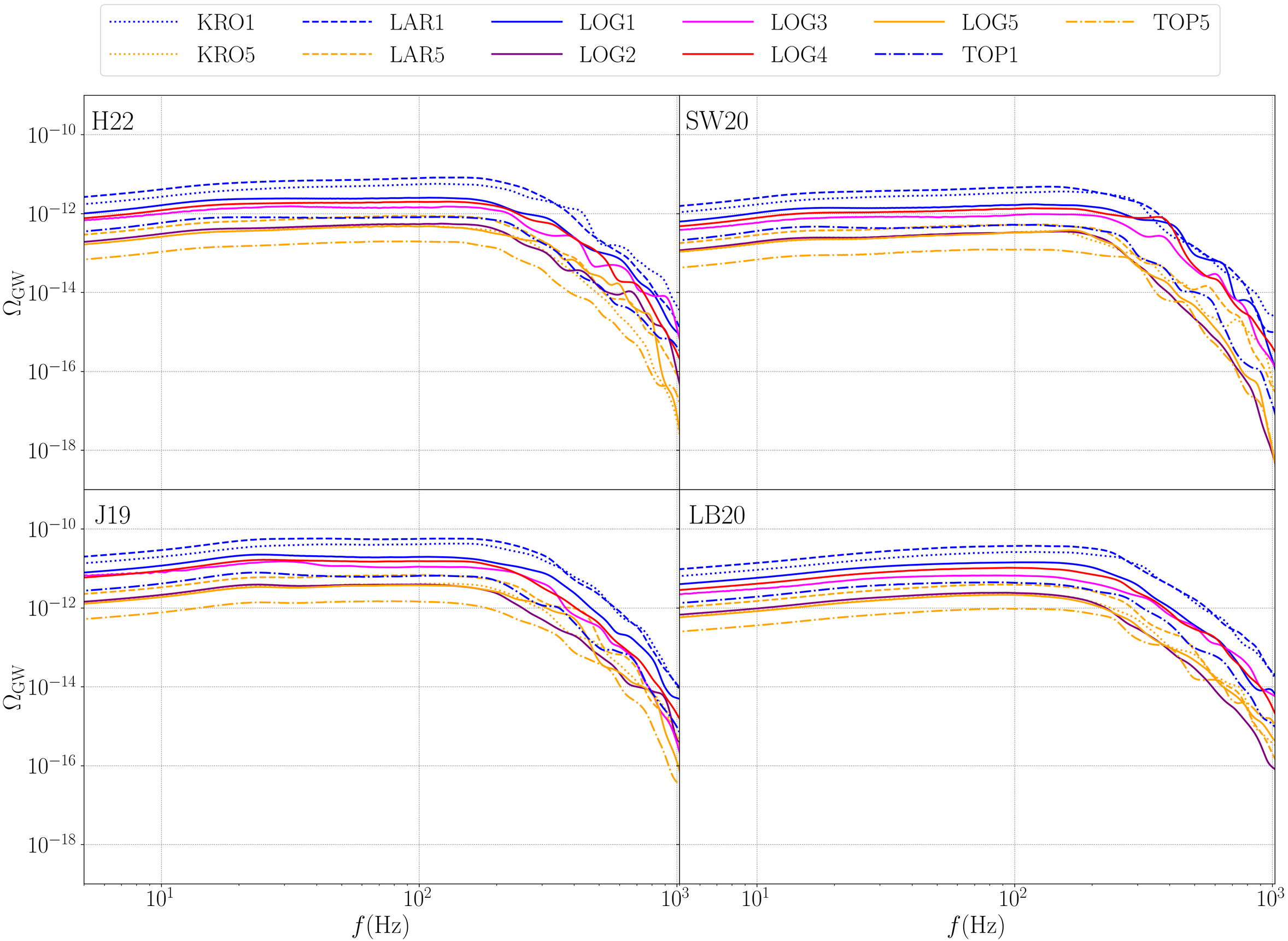}
\caption{The GWB resulting from mergers of BBHs produced by Population III stars, for 4 different SFR models (4 subfigures) and 11 different initial condition models. All models have been taken from~\cite{Filippo_et_al, Costa_et_al}.}
\label{Pop_III_all_models}
\end{figure*}


\section*{Acknowledgments}

We had a number of fruitful discussions on the subject of this paper with
Chris Belczynski, who left us prematurely, and we dedicate this work to his memory. 

This material is based upon work supported by NSF's LIGO Laboratory, which is a major facility fully funded by the National Science Foundation. The authors acknowledge computational resources provided by the LIGO Laboratory and supported by NSF Grants PHY-0757058 and PHY-0823459.

We are indebted to Filippo Santoliquido for many useful discussions on Population III stars. We also thank Ssohrab Borhanian for suggestions on how to run \texttt{GWBENCH}. NK thanks Ansh Gupta, Claire Rigouzzo, Michelle Gurevich, and MS thanks Tania Regimbau for discussions. We thank Nelson Christensen for carefully reviewing this work as a part of the LVK collaboration's internal review process.

NK is supported by King's College London through an NMES Funded Studentship. %
MS acknowledges support from the Science and Technology Facility Council
(STFC), UK, under the research grant ST/X000753/1.

 This manuscript was assigned LIGO-Document number P2400127.  %

\section*{APPENDIX A: DIFFERENCES IN POPULATION III MODELS}
\label{appendix_a}

\subsection*{Initial conditions}

\subsubsection*{Initial mass function}

The initial mass function (IMF) for Population III stars is expected to be quite top-heavy, compared to the one for Population I-II stars~\cite{1998A&A...339..355C, 2002Sci...295...93A, 2004ARA&A..42...79B, 2006ApJ...652....6Y, 2013RPPh...76k2901B, Glover2013, 2022A&A...663A...1G}. This is because the former are characterised by extremely low metallicity and molecular hydrogen is an inefficient coolant, compared to heavier elements~\cite{2004ARA&A..42...79B, 2006MNRAS.369.1437S, 2013MNRAS.433.1094S, 2014ApJ...792...32S, 2014ApJ...781...60H, 2015MNRAS.448..568H, 2020MNRAS.494.1871W, 2021MNRAS.508.4175C, 2021ApJ...910...30T, 2022MNRAS.512..116J, 2022MNRAS.510.4019P, 2023MNRAS.521.5334P}. All considered IMFs favour lower masses, and are essentially power laws~\cite{2013MNRAS.433.1094S, 2014ApJ...792...32S, 2014ApJ...781...60H, 2015MNRAS.448..568H, 2020MNRAS.494.1871W, 2021MNRAS.508.4175C, 2021ApJ...910...30T, 2022MNRAS.512..116J, 2022MNRAS.510.4019P, 2001MNRAS.322..231K, 1998MNRAS.301..569L, 2013MNRAS.433.1094S, 10.1093/mnras/stz1529, 2020MNRAS.495.2475L} (for some models~\cite{1998MNRAS.301..569L, 2013MNRAS.433.1094S, 10.1093/mnras/stz1529, 2020MNRAS.495.2475L} multiplied by an exponential at the lower edge). The most optimistic scenario is given by an IMF which is almost constant~\cite{2013MNRAS.433.1094S, 10.1093/mnras/stz1529, 2020MNRAS.495.2475L}. The aforementioned IMFs were used to sample in primary mass in Ref.~\cite{Costa_et_al}.

\begin{figure}
\centering
\includegraphics[width=\linewidth]{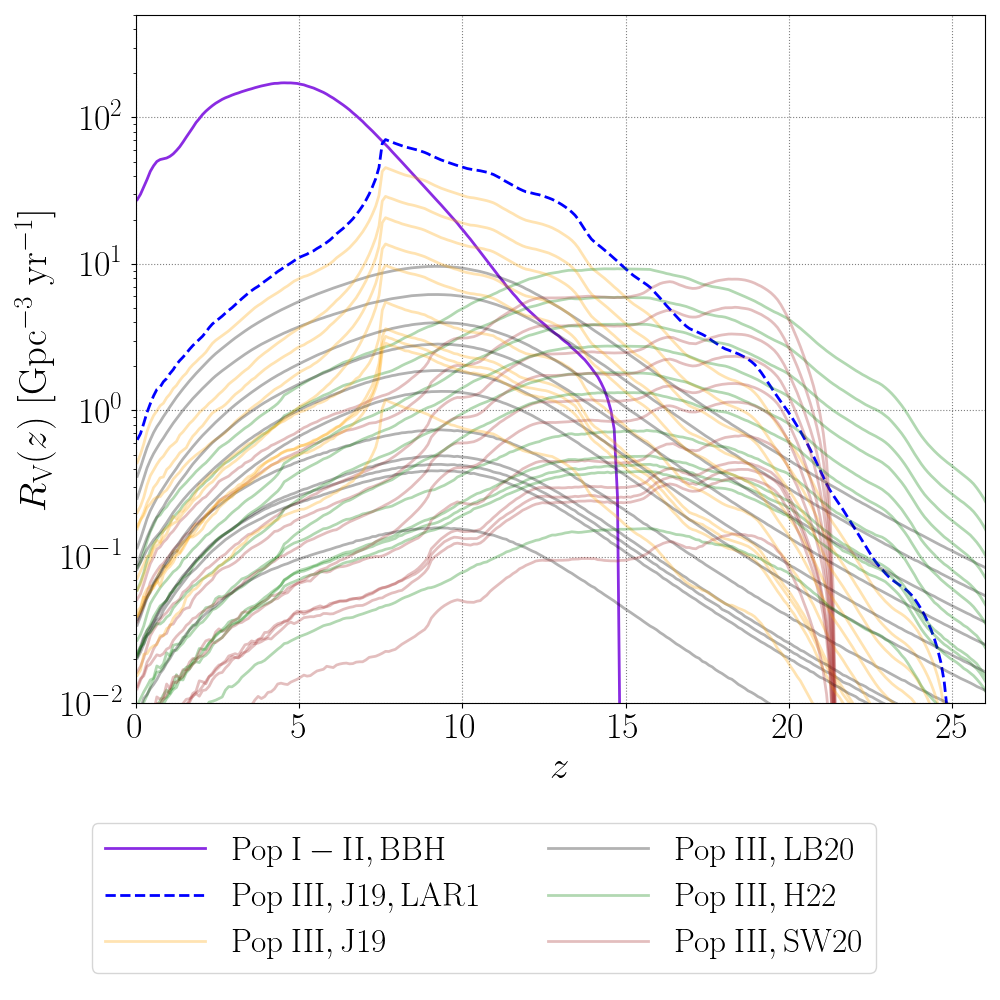}
\caption{The BBH merger rate density (MRD) for Population I-II and the \textit{J19}, \textit{LAR1} Population III model, as computed using \texttt{CosmoRate}. The MRD for every other Population III model is also plotted.}
\label{MRD_peak}
\end{figure}

\subsubsection*{Mass ratio \& secondary mass}

The secondary mass distribution was either identical to the primary mass distribution, or determined assuming a power law distribution~\cite{2012Sci...337..444S, 2013MNRAS.433.1094S} for the mass ratio. In both cases, the final mass ratio distribution also depends on the primary mass distribution.

\subsubsection*{Orbital period}

The orbital period distribution is either a power law favouring shorter periods~\cite{2012Sci...337..444S}, or a Gaussian distribution~\cite{2013MNRAS.433.1094S}.

\subsubsection*{Eccentricity}

The eccentricity distribution is a power law with either positive~\cite{10.1093/mnras/stu1022, 10.1093/mnrasl/slw074, Tanikawa_2021} or negative~\cite{2012Sci...337..444S} exponent, favouring either higher or lower eccentricities, respectively.

\subsection*{Star formation rate}

All considered models for the star formation rate are consistent with the Thomson scattering optical depth value estimated by the
Planck Collaboration~\cite{2016A&A...594A..13P}. Their peak varies significantly, from $z\simeq8$ (\textit{J19}) to $z\simeq20$ (\textit{SW20}), depending on different physical assumptions. First, \texttt{H22} is a semi-analytic model that samples and traces individual stars, based on dark matter halo merger trees and calibrated to reproduce observables in the Universe~\cite{2022ApJ...936...45H, 2023MNRAS.520.3229U}. Next, \textit{J19} was obtained using the hydro-dynamical/$N$-body code \texttt{GIZMO}~\cite{2015MNRAS.450...53H}, considering both the chemical and radiative feedback of core-collapse and pair-instability supernovae. Likewise, \textit{LB20} was the result of simulations with \texttt{GIZMO}, extrapolated to lower redshifts and following a Madau-Dickinson form~\cite{2014ARA&A..52..415M}. Finally, \textit{SW20} was constructed from hydro-dynamical cosmological simulations that ran on the adaptive mesh refinement code \texttt{ENZO}~\cite{2014ApJS..211...19B}.

\begin{figure}
\centering
\includegraphics[width=\linewidth]{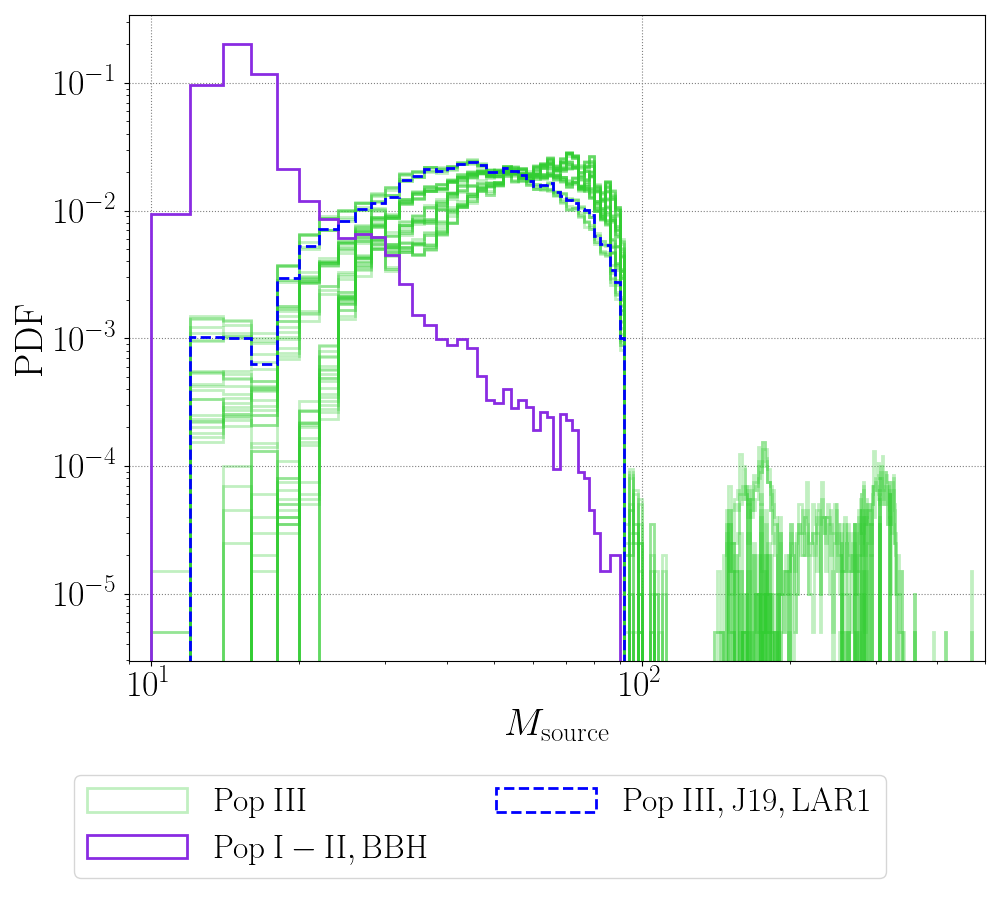}
\caption{The BBH total source mass distribution for Population I-II and the \textit{J19}, \textit{LAR1} Population III model. The latter is clearly characterised by more massive, on average, BBHs. The $M_{\rm source}$ distribution for every other Population III model is also plotted.}
\label{Mass_distribution}
\end{figure}

{\section*{APPENDIX B: ADDITIONAL RESULTS}\label{appendix}

\subsection*{Total background}

We plot in Fig.~\ref{Pop_III_all_models} the $\Omega_{\rm GW}$ from Population III BBH mergers, for the 4 SFR models discussed in Section \ref{pop_models}.
Each of the 4 panels considers a single SFR model (listed at the top left corner), and 11 different IC models. Note that  the individual curves appear in general wobbly in the region $f\geq200$ Hz. This is because high frequencies in the $\Omega_{\rm GW}$ correspond to signals that are emitted by low redshift sources, which are inevitably few in our catalogues. Thus,  $\Omega_{\rm GW}$ at these frequencies is characterised by higher statistical uncertainties. Notice that in all cases there is a plateau, as discussed in Section \ref{total_background}.

\subsection*{Merger rate density}

We show in Fig.~\ref{MRD_peak} the MRD over redshift for Population I-II and Population III BBHs. Even in the most optimistic case (\textit{J19}, \textit{LAR1}), the Population III MRD peak value ($70.6$ Gpc$^{-3}$ yr$^{-1}$) is significantly lower than the one for Population I-II BBH MRD peak value ($172.2$ Gpc$^{-3}$ yr$^{-1}$), and lies at a much higher redshift ($4.55$ and $7.65$, respectively). In all other cases, the MRD peaks at $z\geq7.65$.

\subsection*{Mass distribution}

We plot in Fig.~\ref{Mass_distribution} the total source mass distribution for Population I-II and Population III BBHs. This is expected to be generally higher for the latter because of the corresponding extremely low metallicity, which implies that (a) the initial mass function is more top-heavy as compared to metal-rich stars and (b) there is virtually no mass lost in stellar winds~\cite{2002RvMP...74.1015W, Volpato_2023}. Indeed, the $M_{\rm source}$ distribution takes typically higher values for all Population III models compared to Population I-II. Interestingly, we notice that (\textit{J19}, \textit{LAR1}), which appears rather pessimistic in terms of the expected $M_{\rm source}$, turns out to be the most optimistic case in terms of the $\Omega_{\rm GW}$ (as seen in Fig.~\ref{Pop_III_all_models}) as a result of its high MRD. Note that the fast drop around  $M_{\rm source}\sim100$ can be associated to the pair-instability mass gap, which for Population III BBHs simulated with \texttt{SEVN} has a lower edge at $M_{\rm source}=86M_{\odot}$~\cite{Costa_et_al}.

\bibliography{Pop_III}
\end{document}